\documentclass[a4paper,11pt]{article}
\usepackage[utf8]{inputenc}
\usepackage[english]{babel}
\usepackage{amsmath}
\usepackage{authblk}
\usepackage{graphicx}
\usepackage{url}
\usepackage{float}

\usepackage{cite}

\usepackage{indentfirst}

\usepackage[right=2cm,left=2cm,top=1.75cm,bottom=2.25cm]{geometry}

\title{Fine structure of the diffraction cone: from ISR to the LHC}
\author{D.A. Fagundes}
\affil{\small\it Universidade Federal de Santa Catarina - Campus Blumenau, Rua Pomerode, 710 Salto do Norte, 89065-300 Blumenau, SC, Brazil}
\author{L. Jenkovszky}
\affil{\small\it Bogolyubov Institute for Theoretical Physics, National Academy of Sciences of Ukraine,UA-03680 Kiev, Ukraine}
\author{E.Q. Miranda}
\affil{\small\it InSTEC, Quinta de los Marinos, Ave. Salvador Allende y Lucas, La Habana 10400 Cuba}
\author{G. Pancheri}
\affil{\small\it INFN Frascati National Laboratories, Via E. Fermi 40, 00444, Italy}
\author{P.V.R.G. Silva}
\affil{\small\it Instituto de F\'isica Gleb Wataghin, Universidade Estadual de Campinas, 13083-859, Campinas, SP,  Brazil}

\begin{document}
\maketitle

\begin{abstract}
Following earlier findings, we argue that the low-$|t|$ structure in the elastic diffractive cone, recently reported by the TOTEM Collaboration at $8$ TeV, is a consequence of the threshold singularity required by $t-$channel unitarity, such as revealed earlier at the ISR. By using simple Regge-pole models, we analyze the available data on the $pp$ elastic differential cross section in a wide range of c.m. energies, namely those from ISR to LHC8, obtaining good fits of all datasets. This study hints at the fact that the non-exponential behaviour observed at LHC8 is a recurrence of the low-$|t|$ ``break" phenomenon, observed in the seventies at ISR, being induced by the presence of a two-pion loop singularity in the Pomeron trajectory.   
\end{abstract}

\vskip 0.1cm

$
\begin{array}{ll}
^{1}\mbox{{\small\it e-mail address:}} &
\mbox{daniel.fagundes@ufsc.br}\\
^{2}\mbox{{\it e-mail address:}} &
\mbox{equintana@ucf.edu.cu}\\
^{3}\mbox{{\small\it e-mail address:}} &
   \mbox{jenk@bitp.kiev.ua} \\
^{4}\mbox{{\small\it e-mail address:}} &
  \mbox{Giulia.Pancheri@lnf.infn.it} \\
^{5}\mbox{{\small\it e-mail address:}} &
  \mbox{precchia@ifi.unicamp.br} \\
\end{array}
$
 
\section{Introduction} \label{Intr} 

The recent observation by the TOTEM Collaboration \cite{TOTEM8} of a non purely exponential behaviour of the differential $ pp$ elastic cross-section at  $\sqrt{s}=8$ TeV (LHC8), is discussed in this note in light of two simple Pomeron models \cite{jll_ijmpa_2011} containing a threshold singularity in the Pomeron trajectory, as required by $t$-channel unitarity \cite{Gribov}\footnote{Excellent overviews on the topic can also be found in the Refs. \cite{Fiore2010,KMR2000}}. These models are tested in a wide energy range, namely from ISR to LHC, and in all cases analyzed we find good agreement with the data, in particular in the diffraction cone.


Attempts to relate the “break” seen at the ISR to the square-root singularity in the amplitude (trajectory) go back to 1972 \cite{lnc72}. An empirical model, based on the old Phillips and Barger (PB) model-independent description \cite{Phillips},  has  recently highlighted this relation \cite{Fagundes}. Below we revisit the problem of modelling the elastic differential cross section also in model \cite{Fagundes}, that, however has no well-defined $s$-dependence build in. We shall also test the PB model, appended by a threshold singularity \cite{Fagundes}, with the present LHC8 data \cite{TOTEM8}. 

It is well-known that unitarity constrains the analytic properties of the scattering amplitude. In particular, as shown in Refs. \cite{Barut}, Regge trajectories near the threshold behave as  
\begin{equation} \label{Barut}
\Im \alpha(t)\sim (t-t_0)^{\Re \alpha(t_0)+1/2},
\end{equation}  
where $t_0$ is the lightest threshold, e.g. $t_0=4m_{\pi}^2$ for the $f$ or Pomeron trajectory. 
A good approximation to the lightest threshold is by a square root \cite{lnc72}:
\begin{equation} \label{squareroot}
\alpha(t)\sim \alpha_1\sqrt{t_0-t},
\end{equation}
where $\alpha_1$ is a free parameter, that in Refs. \cite{Prokudin} was associated with the pion mass, $\alpha_1=m_{\pi}/(1\ {\rm GeV}^2)$. While the low-mass $4m_{\pi}^2$  threshold is responsible for the low-$|t|$ structure (the so-called ``break" near $t\approx -0.1$ GeV$^2$), the otherwise exponential shape of the forward cone is provided by the nearly linear behaviour of the Pomeron trajectory beyond the break (in fact, a smooth curvature) and until $t\approx -1$~GeV$^2$. At large $|t|$ the trajectory tends to its logarithmic asymptotics, but this is beyond the scope of the present study (see, {\it e.g.} \cite{jll_ijmpa_2011} and earlier references therein).     

The non-exponential behaviour of the cone at low-$t$ was recently studied in Ref. \cite{Fagundes}, where two options were scrutinized, one with the above threshold singularity in the trajectory and the other one with the form factor, in the line of earlier suggestions by Donnachie and Landshoff (see \cite{DL} and earlier references therein). As already mentioned, recent measurements by the TOTEM Collaboration found deviation from the linear exponential behaviour of the elastic diffraction cone at $8$ TeV \cite{TOTEM8}. We argue that the structure seen at $8$ TeV is a recurrence of that seen earlier at the ISR, both resulting from the threshold singularity imposed by $t$-channel unitarity. To this end we use a simple Regge-pole model extrapolating from the ISR energy region to that of the LHC. In doing so, we use a single ``effective" Regge trajectory that incorporates leading (Pomeron) and non-leading Regge exchanges.    
While in the ISR energy region $f$ exchange is important, it is negligible at the LHC (see e.g. Ref. \cite{jll_ijmpa_2011}).
     
In the present study we use two simple versions of the Regge-pole model incorporating an ``effective" trajectory including the required threshold singularity, with and without a linear term added.  We make a number of fits, one based only on the ISR data to see its extrapolation to the LHC, and the other one including also the new LHC data.

\section{Simple Pomeron Pole (Model 1)} \label{LNC} 
 
The first simple model analysed is the one from Ref. \cite{lnc72}, whose parametrization for the elastic differential cross section follows:
 
 \begin{equation}
  \frac{d\sigma_{el}}{dt} = r\exp\left\{bt + 2 \ln(s/s_0)[\alpha(t)-1]\right\},
 \end{equation}

\noindent where the Pomeron trajectory is written as
  
 \begin{equation}
   \alpha(t) =  \alpha_0 + \alpha' t - \alpha_1\sqrt{t_0-t}
 \end{equation}

 \noindent and $r$ (mbGeV$^{-2}$), $b$ (GeV$^{-2}$), $\alpha_0$ (dimensionless), $\alpha'$ (GeV$^{-2}$), $\alpha_1$ (GeV$^{-1}$) are free parameters, $s_0 = 1$ GeV$^{2}$, $t_0=4m_\pi^2$ (GeV$^{2}$). On performing data analysis with this model we consider two cases (variants): 
\begin{itemize}
 \item[i.]  a linear term is present in the Pomeron (``effective'') trajectory ($\alpha'$ free parameter) - denoted as  \textit{with Linear term};
  \item[ii.] a linear term is absent in $\alpha(t)$ ($\alpha' = 0$ fixed) - denoted as \textit{without Linear term}.
 \end{itemize}

In such case, the Pomeron is regarded as a simple `supercritical' pole in the complex angular momenta plane, with $\Delta = \alpha(0)-1 > 0$ and $\Delta \sim 0.1$.

\section{Double Pomeron Pole (Model 2)} \label{DP}
 
In the framework of the so-called Dipole Pomeron Model (DPM), the Pomeron amplitude is regarded as a two-term amplitude being written as \cite{jll_ijmpa_2011}:

\begin{equation} \label{Eq:DP}
 A_P(s,t) = i \frac{a_P}{b_P}\frac{s}{s_0}[r_1^2(s)e^{r_1^2(s)[\alpha-1]}-\epsilon_Pr_2^2(s)e^{r_2^2(s)[\alpha-1]}],
\end{equation}

\noindent where again 
  
 \begin{equation}
   \alpha(t) =  \alpha_0 + \alpha' t - \alpha_1\sqrt{t_0-t}
 \end{equation}

 \noindent represent the Pomeron (``effective'') trajectory and
 
 \begin{eqnarray}
  r_1^2(s) & = &  b_P + L -i\pi/2,\\
  r_2^2(s) & = & L -i\pi/2,
 \end{eqnarray}

 \noindent $L = \ln(s/s_0)$ and $a_P$, $b_P$, $\alpha_0$ (dimensionless), $\alpha'$ (GeV$^{-2}$), $\alpha_1$ (GeV$^{-1}$) and $\epsilon_P$ are free parameters. In this model, the energy-dependent functions $r_1(s)$ and $r_2(s)$, having logarithm growth in $s$ reflect the unitarization of the Pomeron amplitude at high-energies. Here, $t_0 = 4m_\pi^2$ (GeV$^{2}$) and $s_0 = 1$ GeV$^{2}$ are fixed. Just like with Model 1, we considered here also two possible cases (variants), namely:

\begin{itemize}
 \item[i.] a linear term is present in the Pomeron (``effective'') trajectory ($\alpha'$ free parameter) - denoted as \textit{with Linear term};
  \item[ii.] a linear term is absent in $\alpha(t)$ ($\alpha' = 0$ fixed) - denoted as \textit{without Linear term};
 \end{itemize}

In this framework, from the amplitude in Eq. (\ref{Eq:DP}) we calculate the elastic differential cross section through the following expression: 

\begin{equation}
\frac{d\sigma_{el}}{dt} = \frac{\pi}{s^2}|A_P(s,t)|^2.
\end{equation}

While the parameter $\epsilon_P$ in Eq. (\ref{Eq:DP}) - reflecting absorptive effects - plays a major role in the dip-bump region -  we make the approximation  $\epsilon_P \simeq 0$, namely  to neglect it as we are here primarily  interested in the break at very small $-t$.


\section{Data Analysis and Fits}
In the following we shall briefly discuss our data analysis, presenting in the first place the datasets used in data reductions and then showing the main results achieved.
 
\subsection{Datasets}
 
The datates analyzed here are those of the elastic differential cross section for $pp$ scattering \cite{Cudell_data} at ISR energies, namely in the interval  $23.5 - 62.5$ GeV as well as the recent ones at LHC7, measured by the TOTEM Collaboration \cite{TOTEM7}. In our first approach to data reductions, the LHC8 data \cite{TOTEM8} is not taken into account. However, as we explain along the text, it can be used when properly specified. In table \ref{tab:data} we display the number of points comprising each dataset used for data reductions. As we are analyzing the effect of deviations of exponential behaviour $\sim e^{-b|t|}$ at the diffraction cone, only the data in the interval $|t|: 0.01-0.35$ GeV$^2$ were considered in our estimate and best fit analyses.
 
 
 \begin{table}[H]
  \centering
  \caption{Number of points in each energy in the range of momentum transfer range $|t|: 0.01-0.35$ GeV$^2$ used for data reductions.}
  \vspace*{.2cm}
  \begin{tabular}{cccccccc}\hline
   $\sqrt{s}$ (GeV) & 23.5 & 30.7 & 44.7 & 52.8 & 62.5 & 7000 & 8000\\\hline
   Nº points         &  62  &  88  & 139  &  59  &  53  &  81 & 30\\\hline
  \end{tabular}
  \label{tab:data}
 \end{table}
 
\subsection{Fits - Model 1}
\begin{figure}[H]
  \centering
  \includegraphics*[width=8cm,height=7cm]{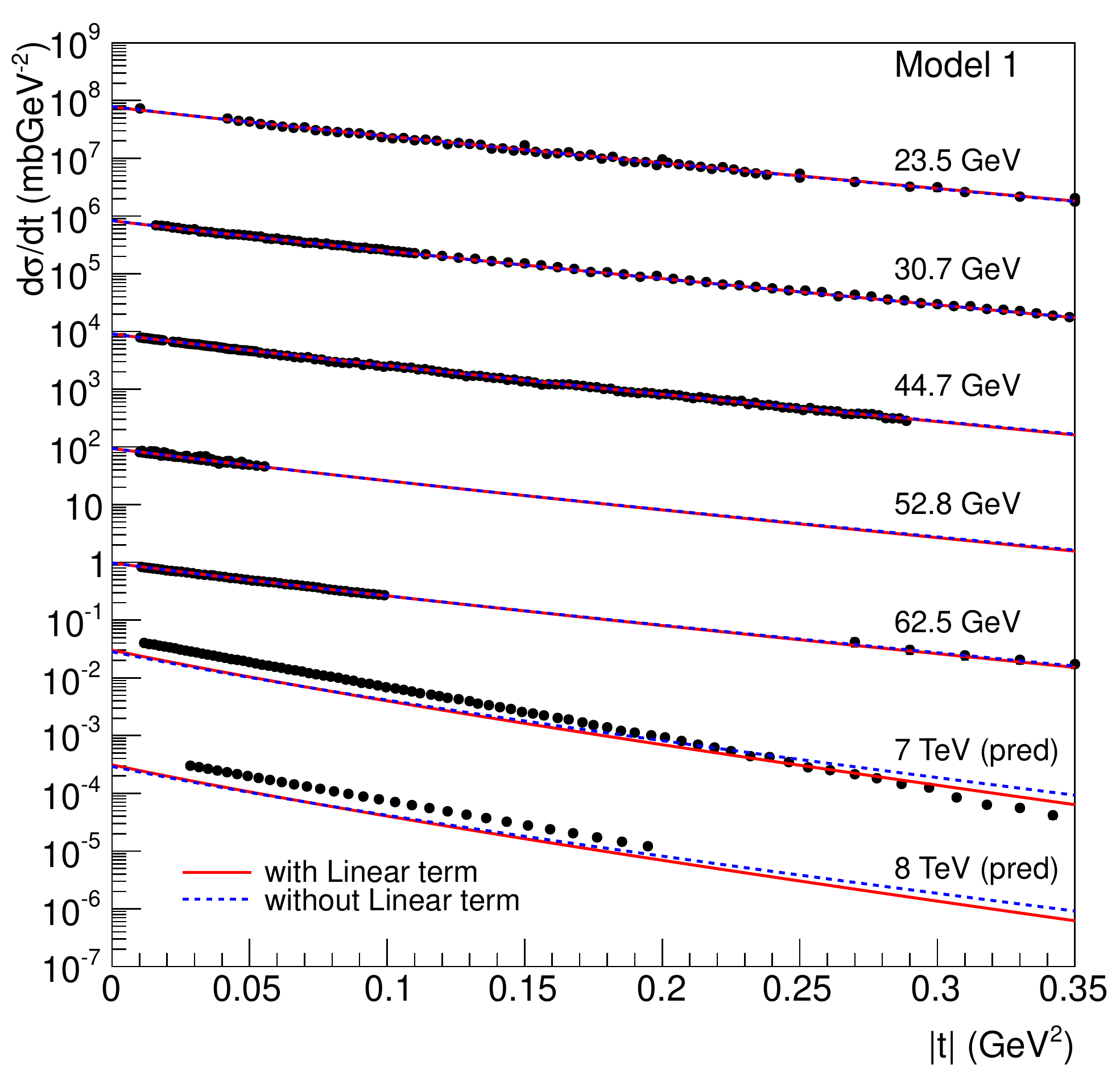}\hspace*{.1cm}
  \includegraphics*[width=8cm,height=7cm]{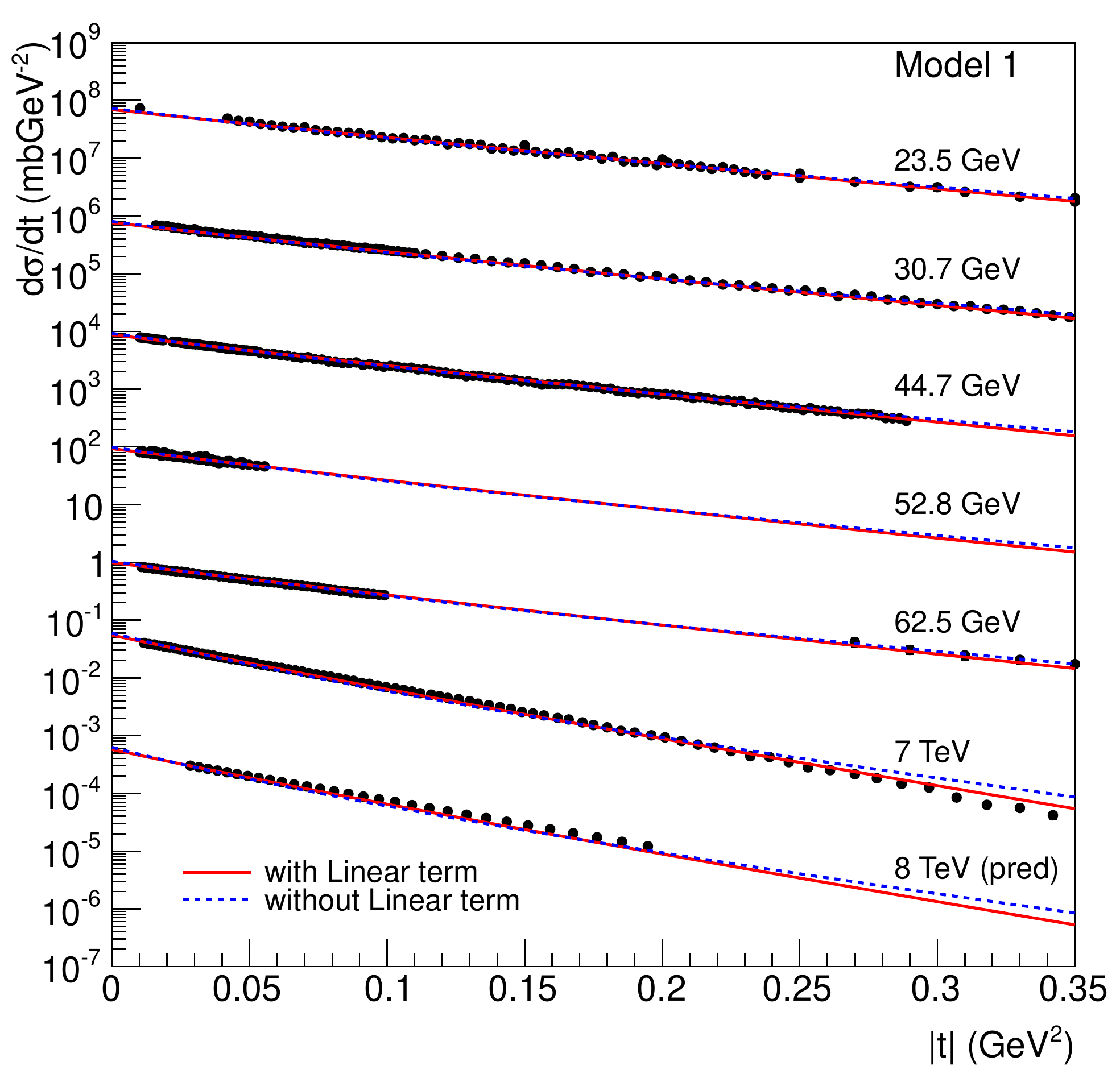}
  \caption{Fits of Model 1,  with or without TOTEM data (\textit{left}) in datasets. In the left panel we show this model prediction for the case where the LHC7 data are not added in the data sets. In the  right panel we refit the data, now including the LHC7 data, as given in \cite{TOTEM7}. In both cases we investigate the effect of a linear term in the Pomeron trajectory, finding a slightly better agreement with data for the model with a linear term. Curves and data are multiplied by $10^{\pm2}$ factors to appear in the same canvas.}
  \label{fig:res_LNC72}
 \end{figure}

\begin{figure}[H]
\centering
\includegraphics[width=12cm,height=12cm]{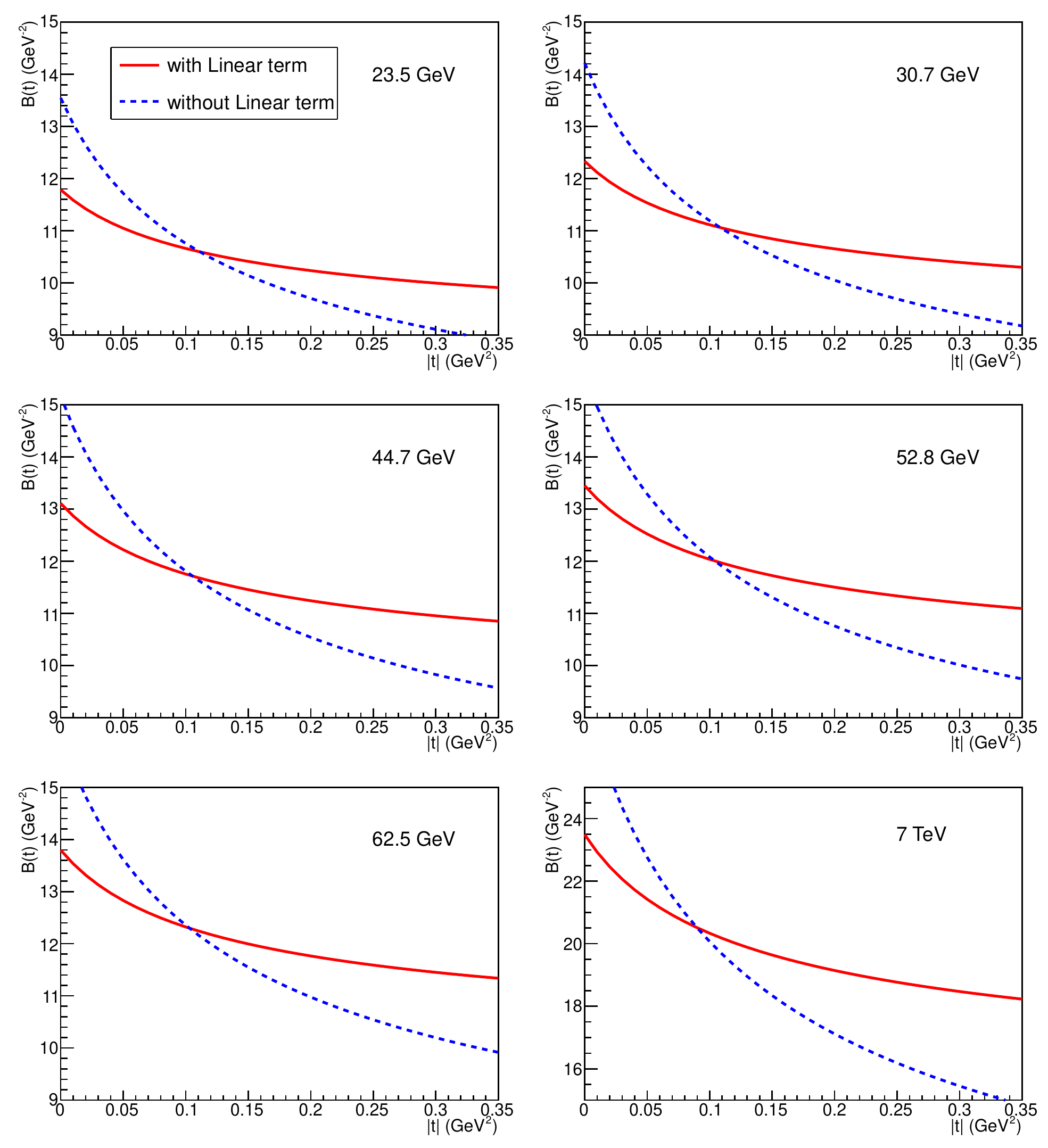}
\caption{Local slope $B(t)$ for the fits of Model 1 (fit with TOTEM data) in two variants: with and without linear term in the trajectory.}
\label{fig:res_B_LNC72_woTOTEM}
\end{figure}

The results obtained with Model 1, shortly described in section \ref{LNC}, are displayed in tables \ref{tab:res_LNC72_woL} and \ref{tab:res_LNC72_wL}, where we consider the two possible cases: with or without linear term. These fits are shown also in figure \ref{fig:res_LNC72}. In figure \ref{fig:res_B_LNC72_woTOTEM} we show the behaviour of the local slope
\begin{equation}
 B(t) = \frac{d}{dt}\left[\ln \frac{d\sigma}{dt}\right]
\end{equation}

\noindent for fits including LHC7-TOTEM data \cite{TOTEM7}.
  
 \begin{table}[H]
  \centering
  \caption{\label{tab:res_LNC72_woL} Fit of Model 1 (with $\alpha'=0$ fixed) with or without the LHC7-TOTEM data added. Statistical information regarding the goodness of fit is also displayed.}
  \vspace*{.2cm}
  \begin{tabular}{c c c}\hline
                &      without TOTEM     & with TOTEM\\\hline
   $r$          &  38.50 $\pm$ 0.72   & 22.76 $\pm$ 0.17 \\
   $b$          &   7.73 $\pm$ 0.11   & 5.594 $\pm$ 0.057 \\
   $\alpha_0$   & 1.1168 $\pm$ 0.0026 & 1.1840 $\pm$ 0.0012 \\
   $\alpha'$    &      0 (fixed)      &    0 (fixed) \\
   $\alpha_1$   & 0.2240 $\pm$ 0.0063 & 0.3399 $\pm$ 0.0027 \\\hline
   $\chi^2/$DOF &         1.77        &        4.97 \\
   DOF          &         397         &        478  \\\hline
  \end{tabular}
 \end{table}

  \begin{table}[H]
  \centering
  \caption{\label{tab:res_LNC72_wL} Fit of Model 1 (with $\alpha'$ free) with or without the LHC7-TOTEM data added. Statistical information regarding the goodness of fit is also displayed.}
  \vspace*{.2cm}
  \begin{tabular}{c c c}\hline
                &      without TOTEM   & with TOTEM\\\hline
   $r$          &  36.410 $\pm$ 0.94   &  22.10 $\pm$ 0.17 \\
   $b$          &   6.91 $\pm$ 0.28    &  5.298 $\pm$ 0.065 \\
   $\alpha_0$   & 1.1140 $\pm$ 0.0028  & 1.1277 $\pm$ 0.0023 \\
   $\alpha'$    &   0.084 $\pm$  0.026 & 0.2597 $\pm$ 0.0091  \\
   $\alpha_1$   & 0.2009 $\pm$ 0.0095  & 0.1371 $\pm$ 0.0075 \\\hline
   $\chi^2/$DOF &         1.76         &        3.26 \\
   DOF          &         396          &        477  \\\hline
  \end{tabular}
 \end{table}

\subsection{Fits - Model 2}

\begin{figure}[H]
  \centering
  \includegraphics*[width=8cm,height=7cm]{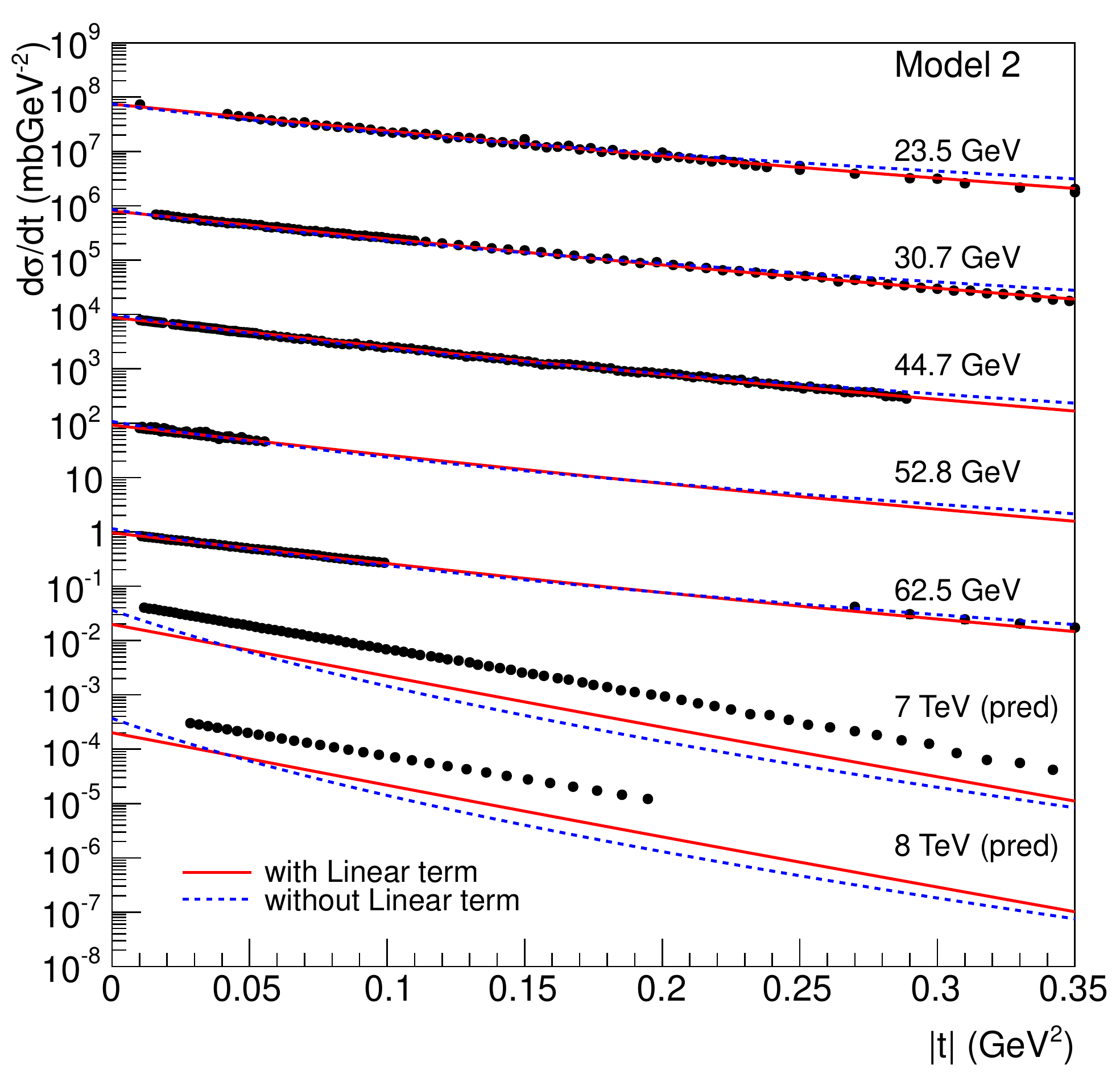}\hspace*{.1cm}
  \includegraphics*[width=8cm,height=7cm]{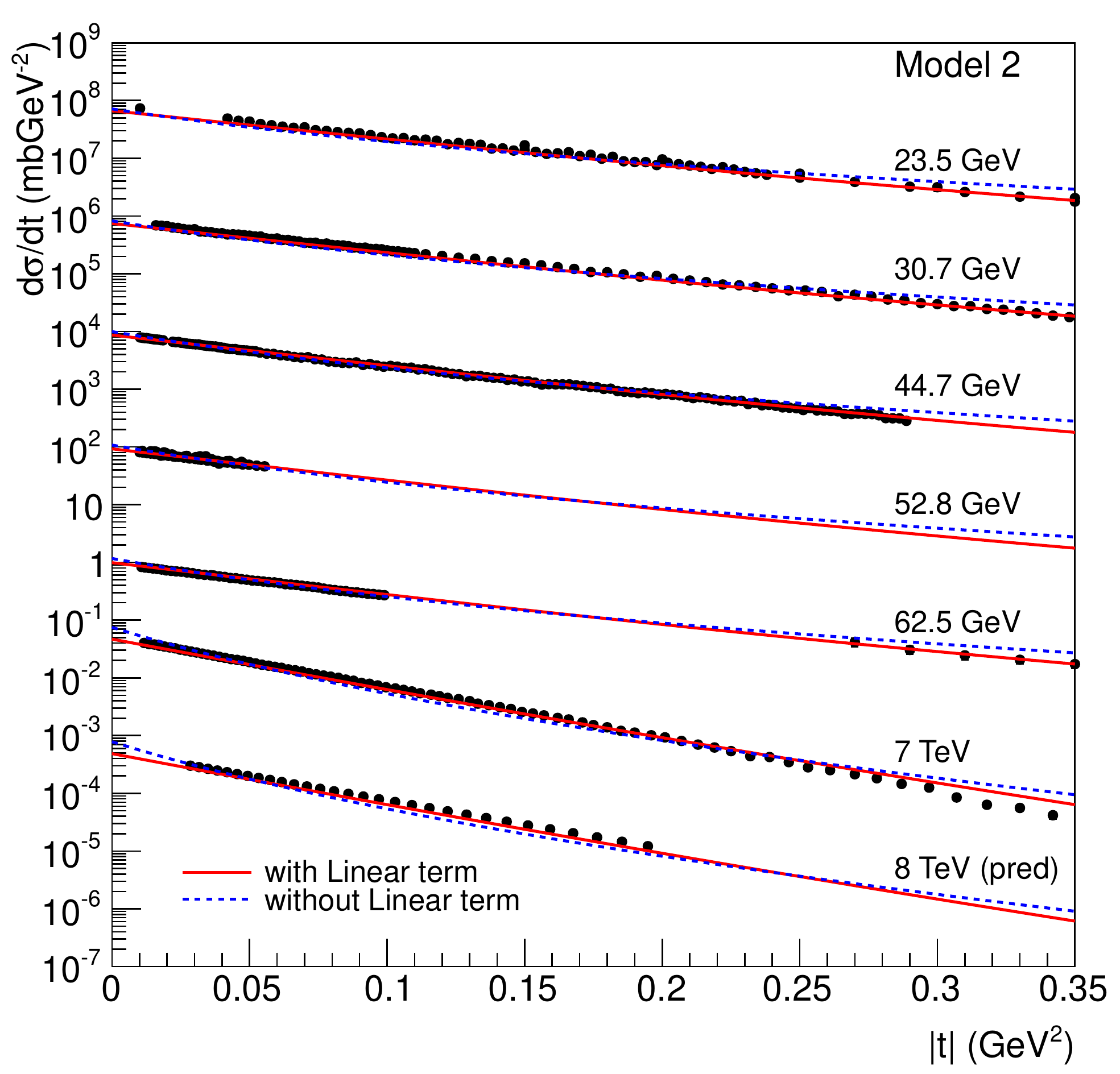}
  \caption{Fits of Model 2,  with or without TOTEM data (\textit{left}) in datasets. In the left panel we show this model prediction for the case where the LHC7 data is not added in the data sets. In the  right panel we refit the data, now including the LHC7 data, as given in \cite{TOTEM7}. In both cases we investigate the effect of a linear term in the Pomeron trajectory, finding a slightly better agreement with data for the model with a linear term. Curves and data are multiplied by $10^{\pm2}$ factors to appear in the same canvas.}
  \label{fig:res_DPM_ePfixed}
 \end{figure}

In this section we present the results obtained with Model 2, namely the Dipole Pomeron Model. As explained in section \ref{DP}, we consider here only fits with $\epsilon_P = 0$ fixed and, as in Model 1, we investigate the effect of the linear term in the trajectory by considering either $\alpha'=0$ fixed or $\alpha'$ as a free parameter. For the first variant, the fits without and with LHC7-TOTEM data are shown in table \ref{tab:res_DPM_woL_ePfixed}. For the latter, the results are shown in table \ref{tab:res_DPM_wL_ePfixed}. For both cases, the comparison among data and curves are given in figure \ref{fig:res_DPM_ePfixed}.

 \begin{table}[H]
  \centering
  \caption{\label{tab:res_DPM_woL_ePfixed} Fits using Model 2 with $\alpha'=0$ fixed and $\epsilon_P=0$ fixed without and with TOTEM data. Statistical informations are also shown.}
  \begin{tabular}{c c c}\hline
                &      without TOTEM   & with TOTEM\\\hline
   $a_P$        &  0.660 $\pm$ 0.046   & 1.270 $\pm$ 0.012 \\
   $b_P$        &   1.230 $\pm$ 0.097  & 5.258 $\pm$ 0.076 \\
   $\alpha_0$   & 1.1483 $\pm$ 0.0013  & 1.15325 $\pm$ 0.00073 \\
   $\alpha'$    &      0 (fixed)       &    0 (fixed) \\
   $\alpha_1$   & 0.6026 $\pm$ 0.0038  & 0.4304 $\pm$ 0.0023 \\
   $\epsilon_P$ &      0 (fixed)       &    0 (fixed) \\\hline
   $\chi^2/$DOF &         14.43        &        19.07 \\
   DOF          &         397          &        478  \\\hline
  \end{tabular}
 \end{table}

 \begin{table}[H]
  \centering
  \caption{\label{tab:res_DPM_wL_ePfixed} Fits using Model 2 with $\alpha'$ free and $\epsilon_P=0$ fixed without and with TOTEM data. Statistical informations are also shown.}
  \begin{tabular}{c c c}\hline
                &      without TOTEM    &   with TOTEM\\\hline
   $a_P$        &    2.91 $\pm$ 0.10    &   1.598 $\pm$ 0.012 \\
   $b_P$        &    6.76 $\pm$ 0.31    &    8.03 $\pm$ 0.11 \\
   $\alpha_0$   &  0.9625 $\pm$ 0.0028  &  1.0260 $\pm$ 0.0015  \\
   $\alpha'$    &  0.6014 $\pm$ 0.0088  &  0.4453 $\pm$ 0.0058    \\
   $\alpha_1$   & -0.0549 $\pm$ 0.0088  & -0.0096 $\pm$ 0.0049 \\
   $\epsilon_P$ &      0 (fixed)        &    0 (fixed) \\\hline
   $\chi^2/$DOF &        1.99           &        5.93 \\
   DOF          &        396            &        477  \\\hline
  \end{tabular}
 \end{table}

 \subsection{Discussion}
 
 In this section we discuss general aspects of the results obtained with Models 1 and 2 for all variants considered.
 
 Considering the fits with and without the linear term in the trajectory, it is clear from figures \ref{fig:res_LNC72} and \ref{fig:res_DPM_ePfixed} that, for both Models, the parametrization with the linear term ($\alpha'$ as a free parameter) describes better the data analyzed, specially when LHC7-TOTEM data is included. It can also be seen in statistical grounds by the values of $\chi^2$ showed in tables \ref{tab:res_LNC72_woL}-\ref{tab:res_DPM_wL_ePfixed}. This effect is more evident for Model 2 (see right panel in fig. \ref{fig:res_DPM_ePfixed} and tables \ref{tab:res_DPM_woL_ePfixed} and \ref{tab:res_DPM_wL_ePfixed}). We conclude that the linear term is necessary to describe the data, even in the low momentum transfer region.
 
 Regarding the extrapolations to 7 and 8 TeV from fits without LHC7-TOTEM, i.e. with only ISR data, from figures \ref{fig:res_LNC72} and \ref{fig:res_DPM_ePfixed} it is evident that our predictions underestimate the data, since we have not used data from $\bar{p}p$ scattering to anchor our fits at very high energies\footnote{In our analysis we do not include the Tevatron’s $\bar pp$ scattering data at 1.8 TeV, where no deviation from exponential behaviour was seen, probably because of their poor statistics.}. For this reason, in the rest of the text, we will focus only in the results obtained with LHC7-TOTEM data.


 \section{Fine Structure in the LHC8-TOTEM data}
  
 If one expects to extrapolate the results here presented to higher energies, specially to 13 TeV, it is of great importance to include the data obtained for the differential cross section in 8 TeV by TOTEM Collaborarion \cite{TOTEM8}. In this section, we present results of these updated fits. 
  
\begin{figure}[H]
  \centering
  \includegraphics*[width=8cm,height=7cm]{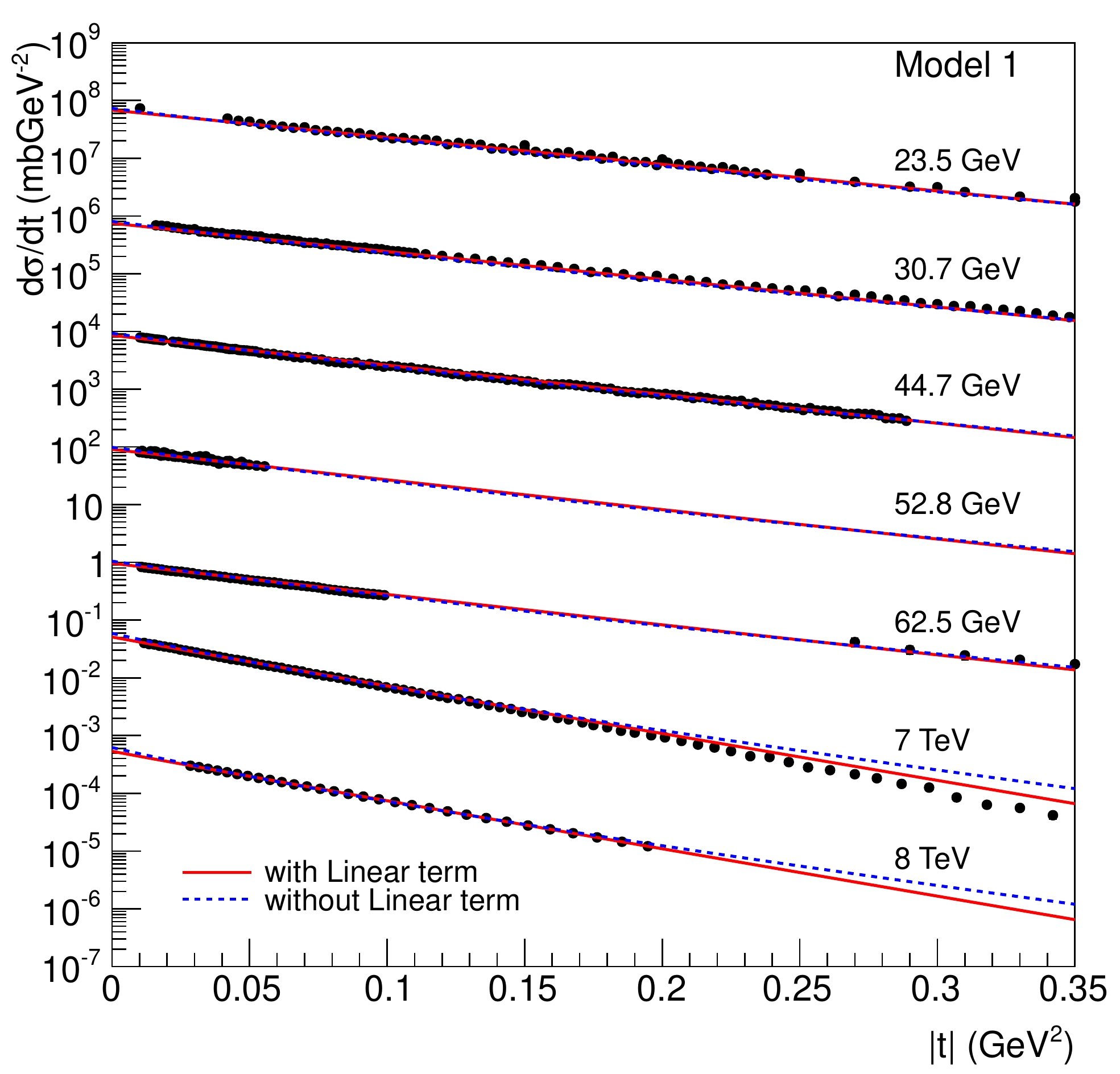}\hspace*{.1cm}
  \includegraphics*[width=8cm,height=7cm]{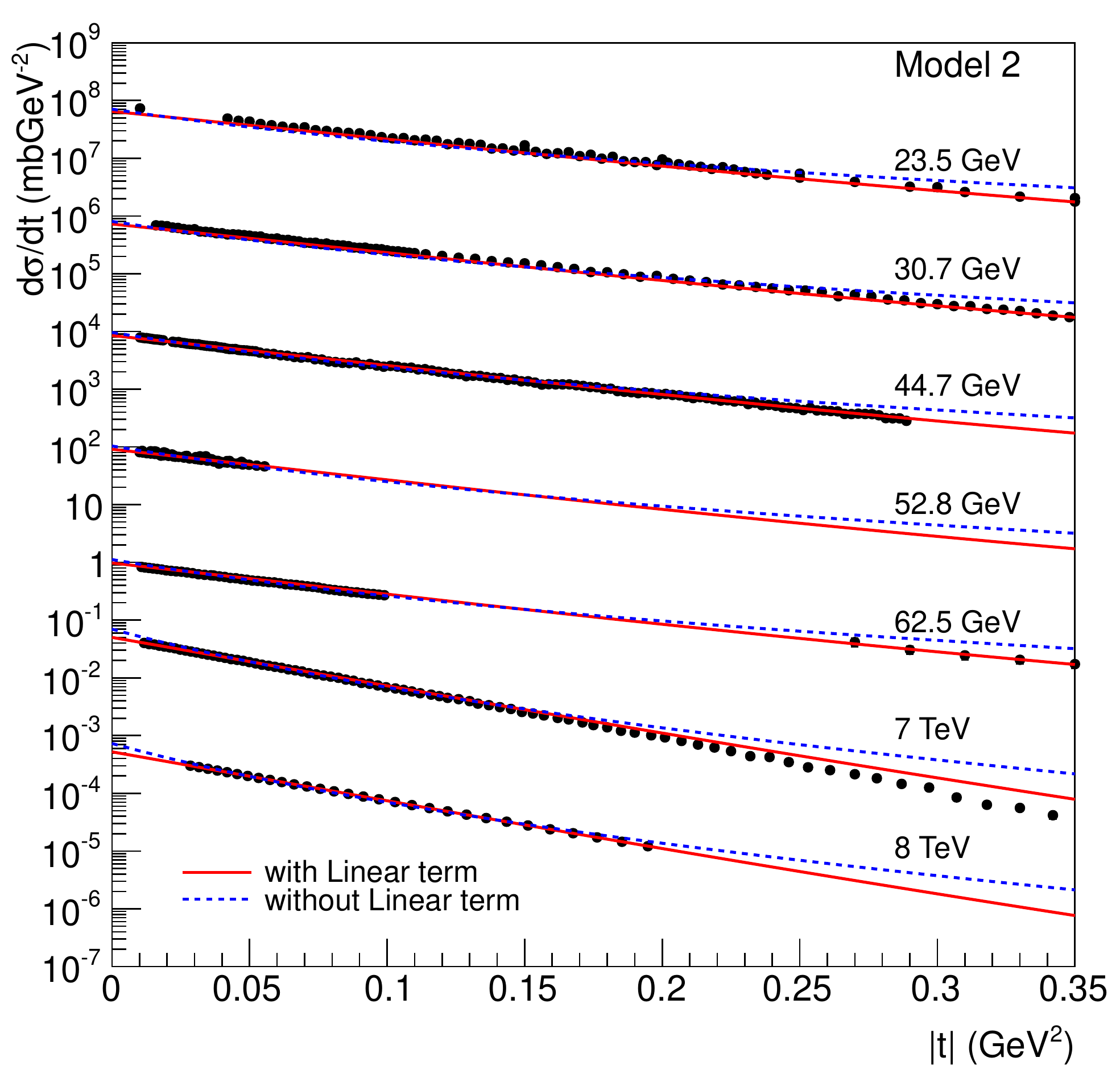}
  \caption{Fit of Model 1 (\textit{left}) and Model 2 (\textit{right}) with and without the linear term in the trajectory with LHC8-TOTEM data \cite{TOTEM8} in datasets. In both cases we investigate the effect of a linear term in the Pomeron trajectory, finding, again, a slightly better agreement with data for the model with a linear term. Curves and data are multiplied by $10^{\pm2}$ factors to appear in the same canvas.}
  \label{fig:res_LNC72_DPM_ePfixed_LHC8}
 \end{figure}

 The results for Model 1 and 2 are showed in tables \ref{tab:res_LNC72_wLHC8_woL_wL} and \ref{tab:res_DPM_wLHC8_woL_wL_ePfixed}, respectively, and in figure \ref{fig:res_LNC72_DPM_ePfixed_LHC8} for both models. For all cases, we consider two variants: with linear term and without linear term in the trajectory.

  \begin{table}[H]
  \centering
  \caption{\label{tab:res_LNC72_wLHC8_woL_wL} Fits using Model 1 including LHC8-TOTEM \cite{TOTEM8} data in two variants: without and with the linear term in the trajectory. Statistical informations are also shown.}
  \begin{tabular}{c c c}\hline
                &   without Linear term      &       with Linear term     \\\hline
   $r$          &  23.008 $\pm$ 0.058    &     22.307 $\pm$ 0.058   \\
   $b$          &   7.193 $\pm$ 0.034    &      6.130 $\pm$ 0.034   \\
   $\alpha_0$   & 1.16548 $\pm$ 0.00030  &    1.09924 $\pm$ 0.00078 \\
   $\alpha'$    &       0 (fixed)        &     0.3195 $\pm$ 0.0035  \\
   $\alpha_1$   & 0.27321 $\pm$ 0.00081  &     0.0402 $\pm$ 0.0027  \\\hline
   $\chi^2/$DOF &        21.4            &           4.51           \\
   DOF          &         508            &            507           \\\hline
  \end{tabular}
 \end{table}

 \begin{table}[H]
  \centering
  \caption{\label{tab:res_DPM_wLHC8_woL_wL_ePfixed} Fits using Model 2 with $\epsilon_P=0$ fixed including LHC8-TOTEM data \cite{TOTEM8} in two variants: without and with the linear term in the trajectory. Statistical informations are also shown.}
  
  \begin{tabular}{c c c}\hline
                 &   without Linear term      &     with Linear term       \\\hline
   $a_P$        &  1.3450 $\pm$ 0.0024   &     1.4716 $\pm$ 0.0023  \\
   $b_P$        &   7.613 $\pm$ 0.045    &      8.604 $\pm$ 0.055   \\
   $\alpha_0$   & 1.13539 $\pm$ 0.00012  &    1.02574 $\pm$ 0.00058 \\
   $\alpha'$    &     0 (fixed)          &     0.4512 $\pm$ 0.0025  \\
   $\alpha_1$   & 0.34239 $\pm$ 0.00063  &     -0.034 $\pm$ 0.0020  \\
   $\epsilon_P$ &     0 (fixed)          &        0 (fixed)         \\\hline
   $\chi^2/$DOF &        81.6            &           7.20           \\
   DOF          &         508            &            507           \\\hline
  \end{tabular}
 \end{table}

 In order to be able to see in more details the deviaton from a pure exponential form, we compare the data and curves to a reference function by means of the ratio:
  
  \begin{equation}
   R = \frac{d\sigma/dt - \mathrm{Ref}}{\mathrm{Ref}},
   \label{eq:ratio_R}
  \end{equation}

 \noindent where $\mathrm{Ref} = A e^{Bt}$ with $A$ and $B$ determined from a fit to the experimental data. For 8 TeV, we have obtained $A = 518.87 \pm 0.40$ mbGeV$^{-2}$ and $B = 19.3880 \pm 0.0088$ GeV$^{-2}$.
  
\begin{figure}[H]
  \centering
  \includegraphics*[width=8cm,height=7cm]{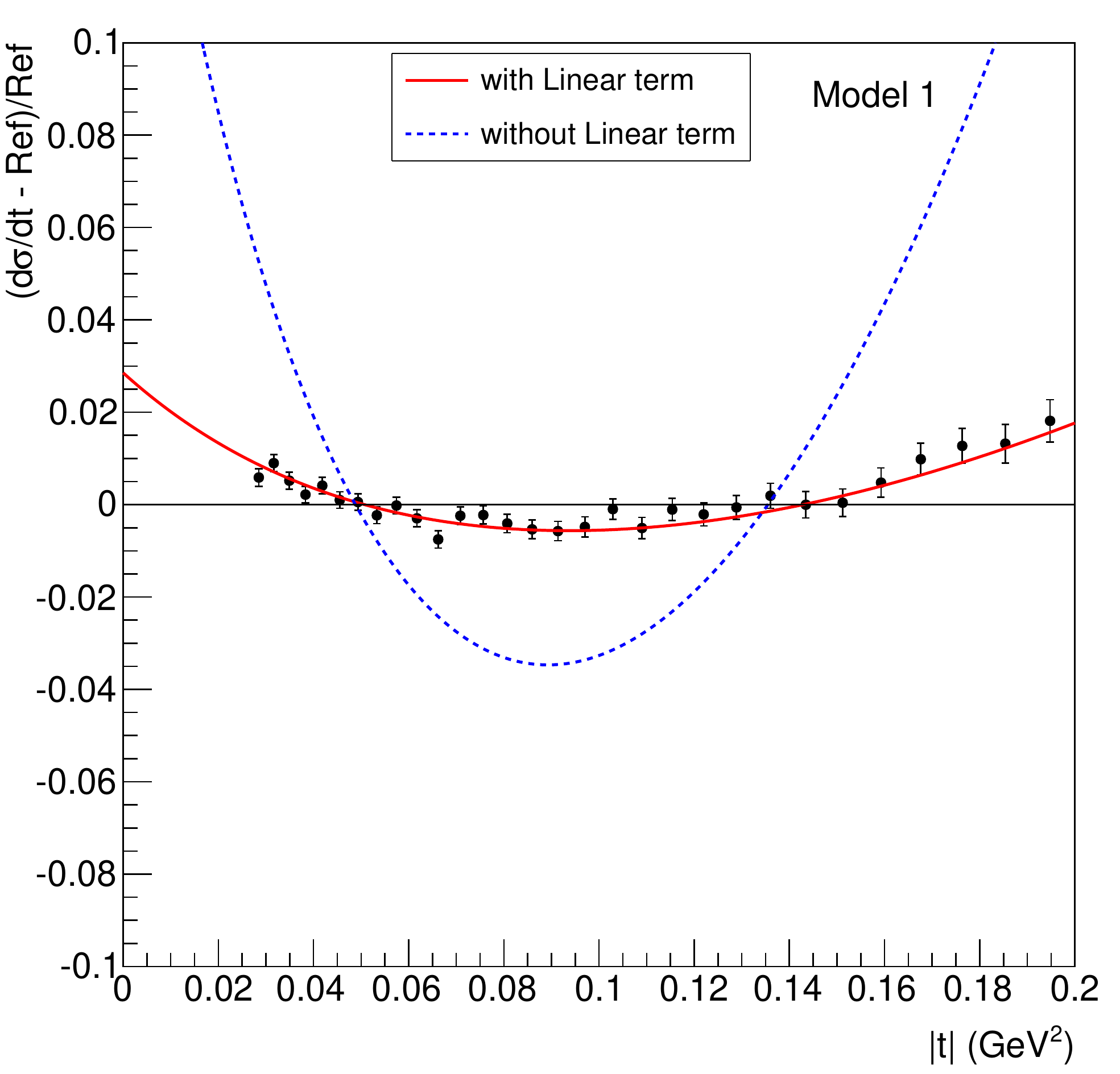}\hspace*{.1cm}
  \includegraphics*[width=8cm,height=7cm]{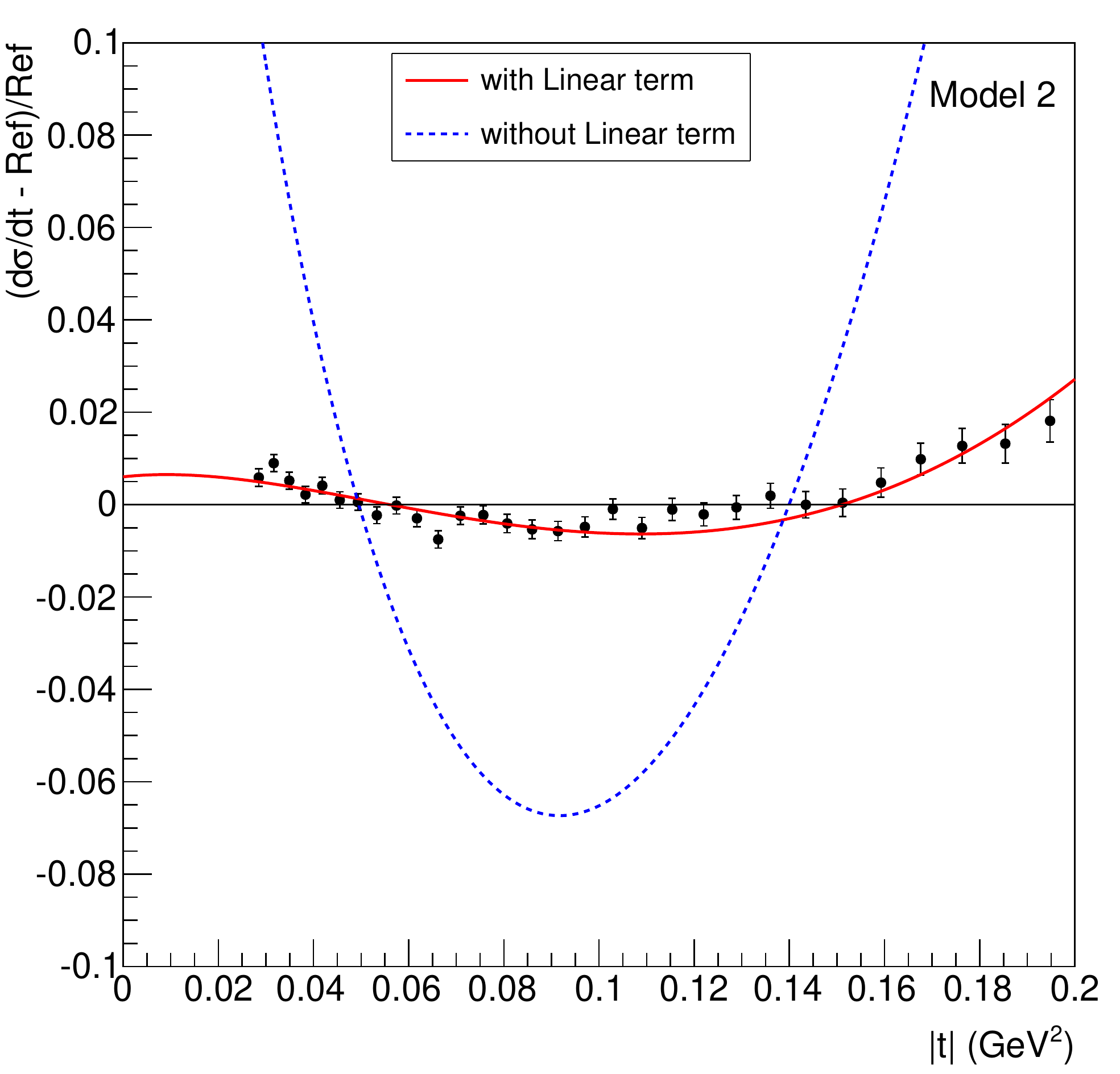}
  \caption{Ratio $R$ (eq.(\ref{eq:ratio_R})) calculated for 8 TeV comparing data and fits with and without linear term for Model 1 (\textit{left}) and Model 2 (\textit{right}).}
  \label{fig:res_LNC72_DPM_LHC8_ref}
 \end{figure}

 In figure \ref{fig:res_LNC72_DPM_LHC8_ref} we compare the results obtained with and without the linear term for the two Models considered. Again, it is clear that the linear term is necessary to describe the data analyzed, particularly the deviation from a pure exponential behaviour presented by 8 TeV data.

\section{The Phillips-Barger model with the pion loop singularity added}
\begin{figure}[h!]
 \centering
 \includegraphics[scale=0.6]{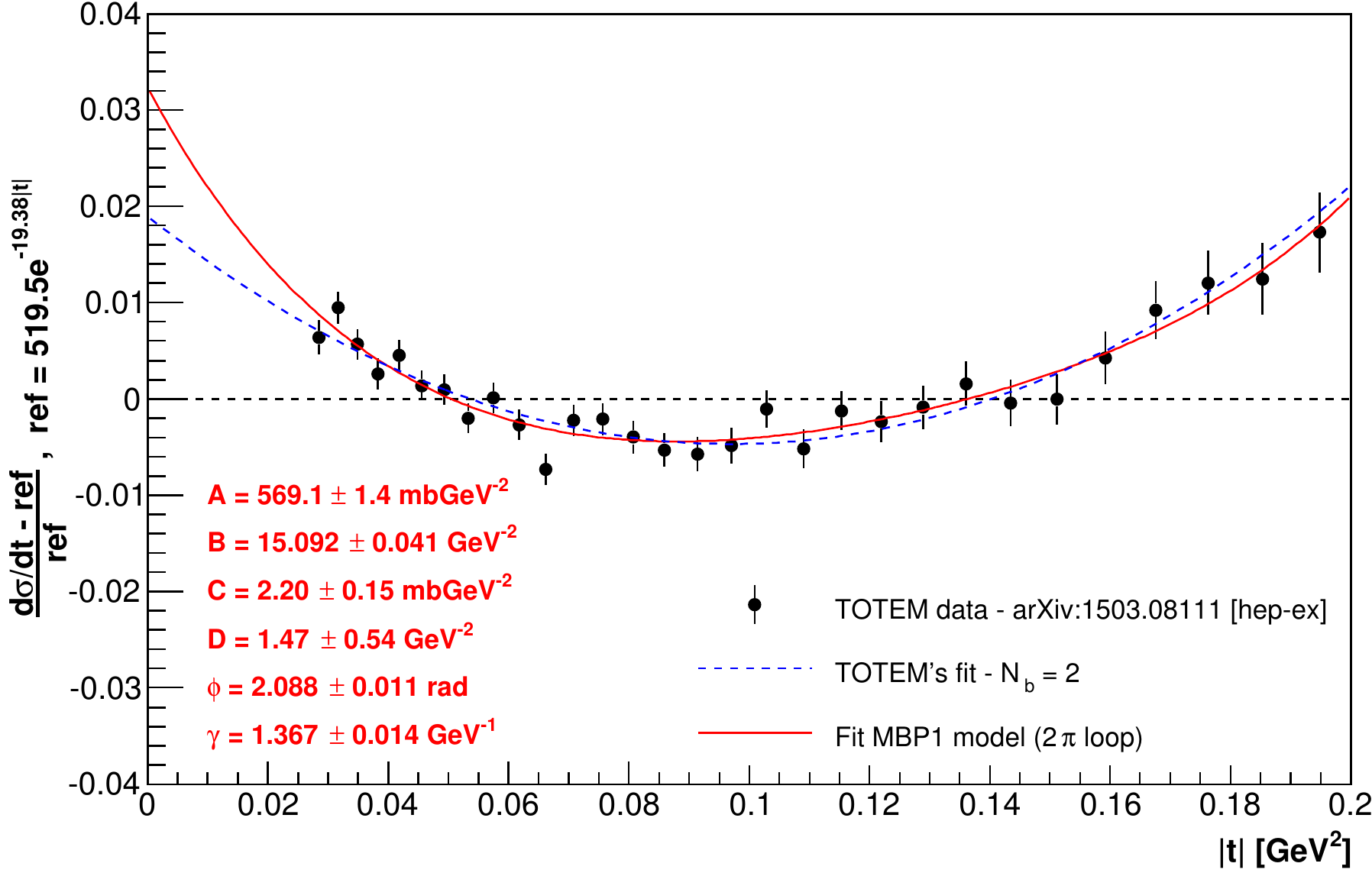}
  \caption{Fit to TOTEM data 
   with the empirical model of \cite{Fagundes} using the pion loop singularity modification of the PB model, MBP1. }
   \label{fig:refBP}
 \end{figure}

Here we show the results from  a model-independent parametrization which can describe both the small and the large $|t|$ region and which had been proposed in 1972 by Phillips and Barger (PB) \cite{Phillips}. This model was recently modified  as \cite{Fagundes}:
\begin{equation}
A_{el}=i \{
F^2_p(t)
\sqrt{A(s)} e^{B(s) t/2} +\sqrt{C(s)}e^{i\phi(s)} e^{D(s) t/2} 
\} \label{eq:mbp}
\end{equation}
and three cases can be considered:
\begin{enumerate}
\item $ F_p(t)=1$ which corresponds to the original PB formulation \cite{Phillips};
\item $F^2_p(t)=e^{-\gamma(s)\sqrt{4 m_\pi^2 -t}}$ which includes a parametrization of the pion loop singularity, labelled as MBP1;
\item $F_p(t)=1/[1+|t|/t_0]^2$ which modifies the very small $-t$ behaviour with a form factor type behaviour, labelled as MBP2.
\end{enumerate}
Of interest to the present discussion is the second case, a square root singularity.
To argue how this model results can be compared with the TOTEM 8 TeV data, we show a fit of the data using the model MBP1, and the comparison with the REF model (one exponential) in fig. ~\ref{fig:refBP}.
We  find that an independent fit, such as the one indicated by the red curve, can give a very good  description of the data, confirming, in this model-independent description,  the agreement with the presence of the pion loop singularity at very small $-t$-values. The parameter values indicated in the figure correspond to option n.2 in Eq.~(\ref
{eq:mbp}) and to a $\chi^2/dof=1.03$.
  
%
  


\section{Conclusions}

We have studied the low-$|t|$ structure of the elastic differential cross section at LHC8, recently reported by TOTEM, using simple Regge-pole amplitudes with a threshold singularity in the Pomeron trajectory. Our results hints at the possibility of interpreting this phenomenon as a recurrence of a similar effect seen at the ISR, in the seventies. In this first approach we focus on the role of the two-pion loop in determining the fine structure of the diffraction cone, namely in affecting the local slope $B(s,t)$. Despite the fact that other sources - such as a form factor, eikonal rescatterings and diffractive dissociation \cite{Ryskin} - can change the behaviour of $B(s,t)$ at small $-t$, our strategy here was to separately investigate the insertion of a two-pion loop in the Pomeron trajectory. Therefore, we have tried to interpolate/extrapolate our fit results for diffraction cone from ISR to LHC. While the interpolation is very good, the extrapolation is only qualitatively satisfactory and shall be investigated in much more detail in a future publication. However, we notice that, the reason of a small mismatch may be attributed mainly to the simplified treatment of 
multiple Regge-pole exchanges replaced by a single ``effective" Regge trajectory. While at the LHC the contribution from secondary trajectories in the nearly forward direction is negligible \cite{jll_ijmpa_2011} this is not true at the ISR, where the $f$ trajectory may contribute by half of the total. Two ways how to cure this insufficiency can be envisaged. One is by including all allowed/required Regge exchanges, namely $\omega, f$ and the Odderon, apart from the Pomeron. The other one is to account for these by introducing energy-dependent parameters in a single ``effective" exchange. 

The fine structure of the cone can be better seen on the local slope $B(s,t)=\frac{d}{dt}\ln\frac{d\sigma}{dt}$. It should be, however, remembered that the result depends on the $t$ interval in which the slope is calculated. The relevant bins can be wide or narrow, overlapping or not. This point was recently discussed in Ref. \cite{JL}. We also notice that simple and elegant formulae extrapolating the forward slope in energy was derived in Ref. \cite{JS}. They read
\begin{equation} \label{JS}
   B(s,t)=k\alpha'(t)\sigma_{tot}(s),\ \ \ B(s_2,t)/B(s_1,t)=\sigma_{tot}(s_2)/\sigma_{tot}(s_1),
\end{equation}
where the coefficient $k$ is determined explicitly in \cite{JS}. The virtue of this formula, following \cite{JS} from 
$s-$channel unitarity is that it is model-independent in the sense that for the total cross section in its r.h.s. on can use either model extrapolations or experimentally measured values of the cross section. The problem with the trajectory is the same as discussed above: in Eq. (\ref{JS}) a single trajectory appears, providing exact predictions when the reaction is dominated by a single Reggeon exchange, as is the case beyond 1 TeV (LHC), dominated by a single Pomeron. Otherwise, an effective trajectory should be used, as discussed above. 

We have used two simple and efficient models of the Pomeron amplitude. The first, presented in Sec. \ref{LNC}, is based on a single pole with a supercritical Pomeron intercept, as advocated by Donnachie and Landshoff \cite{DL}.
The second uses a dipole Pomeron (DP) exchange. Unlike the former, DP promotes logarithmically rising cross sections and consequently the Pomeron is ``softer", its ``supercriticity" $\alpha(0)-1>0$ being about half compared to the case of a simple pole. Also, it contains absorption corrections, regulated by the parameter producing a diffraction minimum and quantified by the parameter $\epsilon_P>0$ in Eq. (\ref{Eq:DP}). In this work, we fixed $\epsilon_P=0$ since we are away from the dip region. This choice improves the fits due to the reduction of the number of free parameters.


The parametrization of trajectories is a key issue in the Regge-pole theory. Linear trajectories are popular for their simplicity, however they contradict unitarity and the analytic properties required by the $S$-matrix theory and the asymptotic constraints, particularly those imposed by the quark model and perturbative quantum chromodynamics (QCD). For practical purposes one chooses a parametrization relevant to the given kinematical region. As shown in the present paper, the economic, parameter-free single square root parametrization, Eq. (\ref{squareroot}), used e.g. in Ref. \cite{Prokudin} produces too strong curvature in the cone, incompatible with the data, as seen in figs. \ref{fig:res_LNC72}, \ref{fig:res_DPM_ePfixed} and \ref{fig:res_LNC72_DPM_ePfixed_LHC8}. The inclusion of a linear term balances this distortion providing good fits to the data.

Finally we note that the reason of non-observation of any fine structure in the cone at the Tevatron or RHIC may be attributed to poor statistics in relevant experiments. As already mentioned, that is one the reasons we did not perform fits using $\bar{p}p$ data. It is also possible that for the same reason it was not observed in diffraction dissociation either. Future experiments may reveal similar effects at the very small $-t$ domain of high-energy diffractive scattering process, either in elastic or inelastic processes. Investigations on this subject are currently in progress.

\section*{Acknowledgements}
 
P.V.R.G.S., E.M. and L.J. thank the International Institute of Physics (ITP) at Natal, Brazil, where this work started, for its hospitality and support during the \textit{School and Workshop on New Trends in High-Energy Physics} in October-November, 2014. Research supported by FAPESP, Contract 2013/27060-3 (P.V.R.G.S.) is acknowledged. L.J. was supported also by DOMUS of the Hungarian Academy of Sciences.


\begin{thebibliography}{}
 
  
\bibitem{TOTEM8} G. Antchev \textit{et al.} (TOTEM Collaboration), Nucl. Phys. B \textbf{899} (2015) 527-546, arXiv:1503.08111 [hep-ex].

\bibitem{jll_ijmpa_2011} L.L. Jenkovszky, A.I. Lengyel, D.I. Lontkovszkyi, Int. J. Mod. Phys. A \textbf{26} (2011) 4755-4771.
 
\bibitem{Gribov} A.A. Anselm and V.N. Gribov, Phys. Lett. B \textbf{40} (1972) 487.

\bibitem{KMR2000} V.A. Khoze, A.D. Martin and M.G. Ryskin, Eur. Phys. J. C \textbf{18} (2000) 167-179. 

\bibitem{Fiore2010} R. Fiore et al, Int. J. Mod. Phys. A \textbf{24} (2009) 2551-2599.

\bibitem{lnc72} G. Cohen-Tannoudji et al, Lett. Nuov. Cim. \textbf{5}, 957 (1972).

\bibitem{Phillips} R.J.N. Phillips and V.N. Barger, Phys. Lett. B \textbf{46} (1973) 412-414. 

\bibitem{Fagundes} D.A.~Fagundes et al, Phys.\ Rev.\ D  \textbf{88} (2013) 094019.


\bibitem{Barut} A.O. Barut and D.E. Zwanziger, Phys. Rev. \textbf{127} (1962) 974;\\
\hspace*{.1cm} V.N. Gribov and Ya.I. Pomeranchuk, Nucl. Phys. \textbf{38} (1962) 516; \\
\hspace*{.1cm} R. Oehme, at the \textit{Strong interactions and high energy physics}, Scottish Summer School 1963, ed. R.G. Moorhouse et al., Edinburgh (1964), p. 129.

\bibitem{Prokudin} R. Fiore, L.L. Jenkovszky, F. Paccanoni and A. Prokudin, Phys. Rev. D {\bf 68}, 013505 (2003); hep-ph/0302195.  

\bibitem{DL} P.V. Landshoff, {\it Pomeron physics: and updade}, hep-ph/0010315; P.V.~Landshoff, {\it Total cross sections at the LHC}, hep-ph/0709.0395.
  
\bibitem{Cudell_data} http://www.theo.phys.ulg.ac.be/alldata-v2.zip.

\bibitem{TOTEM7} G. Antchev \textit{et al.} (TOTEM Collaboration), Europhys. Lett. \textbf{101} (2013) 21002.
  
\bibitem{Ryskin} V.A. Khoze, A.D. Martin and M.G. Ryskin,  J. Phys. G \textbf{42} (2015) 2, 025003.

  
\bibitem{JL} Laszlo Jenkovszky and Alexander Lengyel, Acta Phys. Pol. B \textbf{46}, 863 (2015), arXiv:1410.4106 [hep-ph]. 
  
\bibitem{JS} L.L. Jenkovszky, B.V. Struminsky, Yadernaya Fizika {\bf 39} 1251 (1984) (Engl. translation: Sov. J. Nucl. Phys.).
  
   
  \end{thebibliography}
\end{document}